\documentclass[12pt, letterpaper]{revtex4-1}
\usepackage{graphicx}
\usepackage{amsmath,amssymb}
\usepackage{subfig}
\usepackage{overpic}
\begin{document}
\newpage
\textit{Accepted for publication in the Special issue of Pamir2016 in the journal Magnetohydrodynamics}

\title{\uppercase{Azimuthal Magnetorotational Instability at low and high magnetic Prandtl numbers}}

\author{A. Guseva$^{1,2}$,  R. Hollerbach$^3$, A. P. Willis$^4$, M. Avila$^{1,2}$}

\address{$^1$ Center of Applied Space Technology and Microgravity (ZARM), University of Bremen,\\
 Am Fallturm, 28359 Bremen, Germany }
\address{$^2$ Institute of Fluid Mechanics, Friedrich-Alexander University of Erlangen-N{\"u}rnberg, Cauerstra{\ss}e 4, 91058 Erlangen, Germany}
\address{$^3$ School of Mathematics, University of Leeds, Leeds LS2 9JT, UK}
\address{$^4$ School of Mathematics and Statistics, University of Sheffield, Sheffield S3 7RH, UK}

\vspace{10pt}

\begin{abstract}
The magnetorotational instability (MRI) is considered to be one of the most powerful sources of turbulence in hydrodynamically stable quasi-Keplerian flows, such as those governing accretion disk flows. Although the linear stability of these flows with applied external magnetic field  has been studied for decades, the influence of the instability on the outward angular momentum transport, necessary for the accretion of the disk, is still not well known. In this work we model Keplerian rotation with Taylor-Couette flow and imposed azimuthal magnetic field using both linear and nonlinear approaches. We present scalings of instability with Hartmann and Reynolds numbers via linear analysis and direct numerical simulations (DNS) for the two magnetic Prandtl numbers of $1.4 \cdot 10^{-6}$ and $1$. Inside of the instability domains modes with different axial wavenumbers dominate, resulting in sub-domains of instabilities, which appear different for each $Pm$. The DNS show the emergence of 1- and 2-frequency spatio-temporally oscillating structures for $Pm=1$ close the onset of instability, as well as significant enhancement of angular momentum transport for $Pm=1$ as compared to $Pm=1.4 \cdot 10^{-6}$.
\end{abstract}

\maketitle



\section{Introduction}
\label{sec:intro}
The problem of the magnetorotational instability is closely connected with the accretion disk problem. These huge astrophysical objects, consisting of dust and ionized gas, possess an amount of angular momentum which prevents them from contraction. The angular momentum has to be somehow extracted and transported outwards, and the most effective way to do that is turbulence. Accretion disk flows have, however, hydrodynamically stable Keplerian angular velocity profile $\Omega \sim r^{-3/2}$. One way to destabilize such profiles comes from magnetic field action \cite{velikhov1959, balbus1991}, which may arise from the dynamo in the disk or from the central object (e.g.\ a star). Here we model the accretion disk flow with the Taylor-Couette setup, consisting of two co-rotating cylinders and conducting fluid between them, with  applied azimuthal magnetic field.  The rotation of the cylinders was fixed so that angular momentum increases but angular velocity decreases (quasi-Keplerian flow). In the absence of magnetic field angular momentum transport in this linearly stable flow is supported only by molecular effects (viscosity). In the presence of external azimuthal magnetic field flow is destabilized via the so-called azimuthal magnetorotational instability (AMRI), as shown by linear analysis of Hollerbach et al.\ \cite{hollerbach2010}. R{\"u}diger et al.\ \cite{ruediger2015} studied the AMRI transport in Taylor-Couette system with direct numerical simulations for large magnetic Prandtl numbers ($0.1 < Pm < 1$) and moderate Reynolds numbers $Re < 2 \cdot 10^3$. They found that the resulting turbulence  contributes to the total angular momentum transport and they suggested that the turbulent viscosity scales as $\nu_t/\nu \propto  \sqrt{Pm} Re$.
In this work we compare the dynamics of the system at large $Pm=1$ and small $Pm=1.4 \cdot 10^{-6}$ of InGaSn alloy, exploring wide range of $Re$ ($Re < 6 \cdot 10^3$ for large and $Re< 4 \cdot 10^4$ for small $Pm$).  First, we perform a  linear stability analysis of the flow using the method of Hollerbach et al.\ \cite{hollerbach2010} in order to define scalings of the instability borders and parameter paths of the maximum growth rates of perturbations. Second, by performing direct numerical simulations we trace  the transition to turbulence in the system. Finally we estimate angular momentum transport with our nonlinear pseudo-spectral DNS method, described in \cite{guseva2015}, which allows us to investigate both high and low $Pm$ with good accuracy. 

\section{Model}
\label{sec:presentation}
We consider an incompressible viscous conducting fluid that is sheared between two rotating cylinders. The velocity and magnetic field are determined by the MHD equations:
\begin{equation}
\label{eq:NSeq}
   (\partial_{t} + \mathbf{u}\cdot\nabla) \mathbf{u} = -\nabla p + \Delta \mathbf{u} + \frac{Ha^2}{Pm} (\nabla \times \mathbf{B}) \times \mathbf{B},
\end{equation}
\begin{equation}
\label{eq:Ieq}
( \partial_{t} - \frac{1}{Pm} \Delta) \mathbf{B}  = \nabla \times (\mathbf{u} \times \mathbf{B}),
\end{equation}
where $Ha={B_0}\delta(\sigma/\rho\nu)^{1/2}$ is the Hartmann number, $Pm=\nu/\lambda$ the magnetic Prandtl number ($\rho$ - density, $\nu$ - kinematic viscosity, $\lambda$ - magnetic diffusivity, $\sigma$ - electrical conductivity).  These equations were non-dimensionalized with the following scales: the gap between cylinders $\delta=r_o-r_i$, viscous time scale $\delta^2/\nu$, external magnetic field of magnitude $B_0$. Reynolds number was defined with rotation of inner cylinder and the gap $\delta$: $Re=\Omega_i r_i \delta /\nu$. It does not appear explicitly in the equations, but it appears in the boundary conditions on the cylinders (no-slip for velocity, insulating for magnetic field). In the axial direction periodic boundary conditions were used. The rotation rate $\mu=\Omega_o/\Omega_i$ together with the radius ratio $\eta=r_i/r_o$ defines the rotation regime. In the following we fix  $\mu=0.26$ and $\eta=0.5$ resulting in a hydrodynamically stable flow in the absence of magnetic field (according to the Rayleigh criterion for stability $\eta^2 < \mu$ \cite{rayleigh1917}). The imposed magnetic field $\mathbf{B_0}=B_0(r_i/r) \mathbf{\hat{e}_{\phi}}$ is directed azimuthally. 

\section{Linear stability of the flow}
\label{sec:lin}

By linearizing the nonlinear equations (\ref{eq:NSeq}-\ref{eq:Ieq}) about the basic flow:
\begin{equation}\label{eq:couette}
   V(r) = \frac{Re}{1+\delta}
   \left[
      ( \frac{\mu}{\delta} - \delta ) r + \frac{\delta}{(1-\delta)^2}
      ( 1 - \mu ) \frac1{r}
   \right], 
\end{equation}
and imposed magnetic field 
 \begin{equation}\label{eq:mf}
\mathbf{B_0}=B_0 (r_i/r) \mathbf{\hat{e}_{\phi}},
\end{equation}
and considering disturbances in the form of
 \begin{equation}\label{eq:disturb}
u' \sim \mathrm{exp} (i m \phi + i k z + \gamma t),
\end{equation}
 we can find the parameter regions where the real part of the growth rate $\Re [ \gamma ]$ is positive, and thus perturbations grow.  The details of the method were described in
[3]. In the case of AMRI  the most unstable eigenmode is nonaxisymmetric ($m=1$ for radius ratio  $\eta=0.5$). The axial wavenumber $k$ is also optimized so the most unstable mode is obtained. 
 
\begin{figure}[h!]
\centering
\subfloat[]{
\begin{overpic}[width=0.45\textwidth]{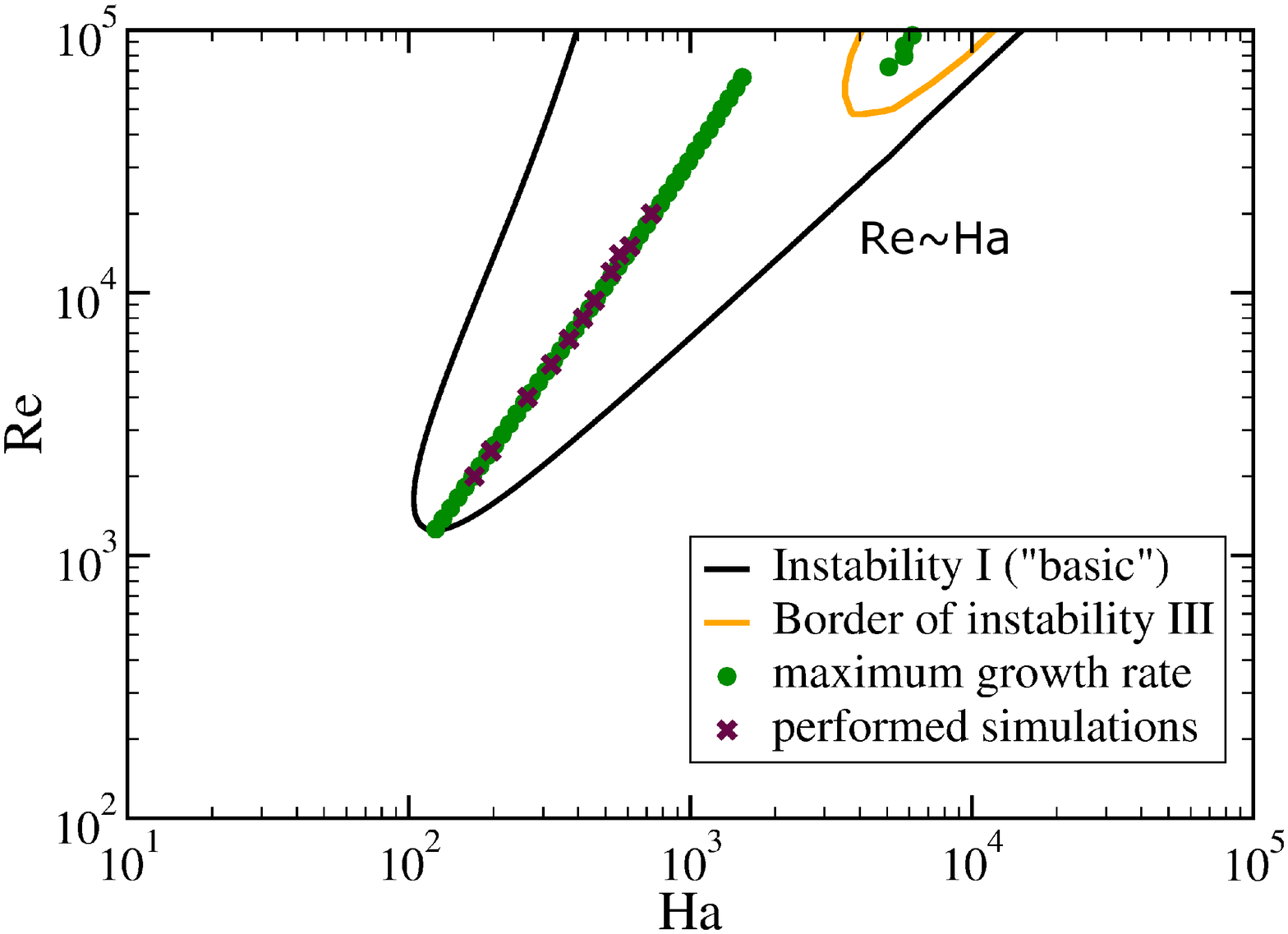}
\put(50,75){\normalsize (a)}
\end{overpic}
}
\subfloat[]{
\begin{overpic}[width=0.45\textwidth]{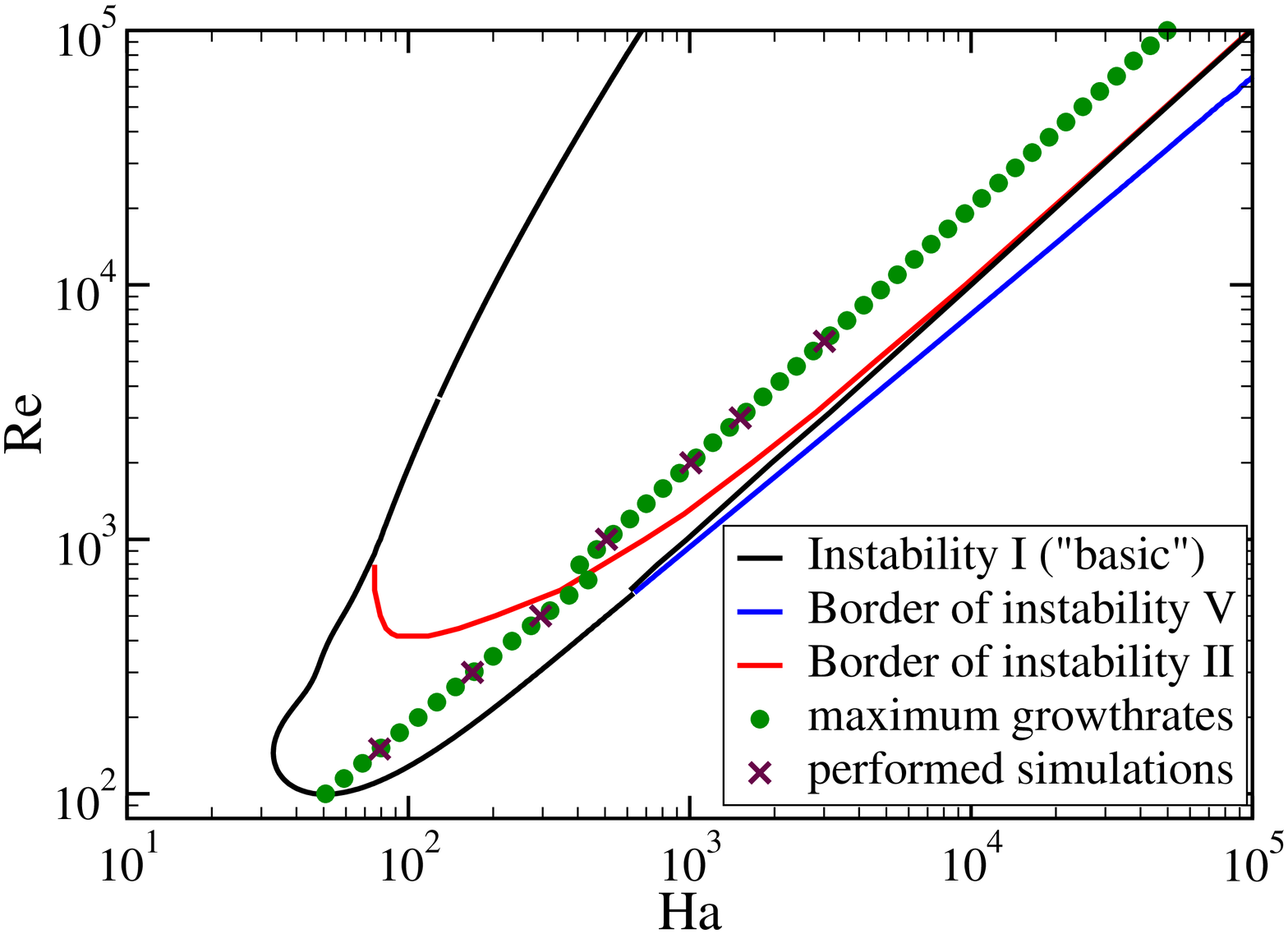}
\put(50,75){\normalsize (b)}
\end{overpic}
}
\caption{Stability maps for $\mu=0.26$ and different $Pm$: ($a$) $Pm=1.4\cdot10^{-6}$, ($b$) $Pm=1$. Yellow, red and blue line show the locations in the $Ha-Re$ plane where axial wavenumber changes discontinuously; green dots correspond to the maximum growth rates.}
\label{Fig:stabmaps}
\end{figure}

\begin{figure}[h!]
\centering
\subfloat[]{
\begin{overpic}[width=0.45\textwidth]{Re_60000_k.eps}
\put(50,75){\normalsize (a)}
\end{overpic}
}
\subfloat[]{
\begin{overpic}[width=0.45\textwidth]{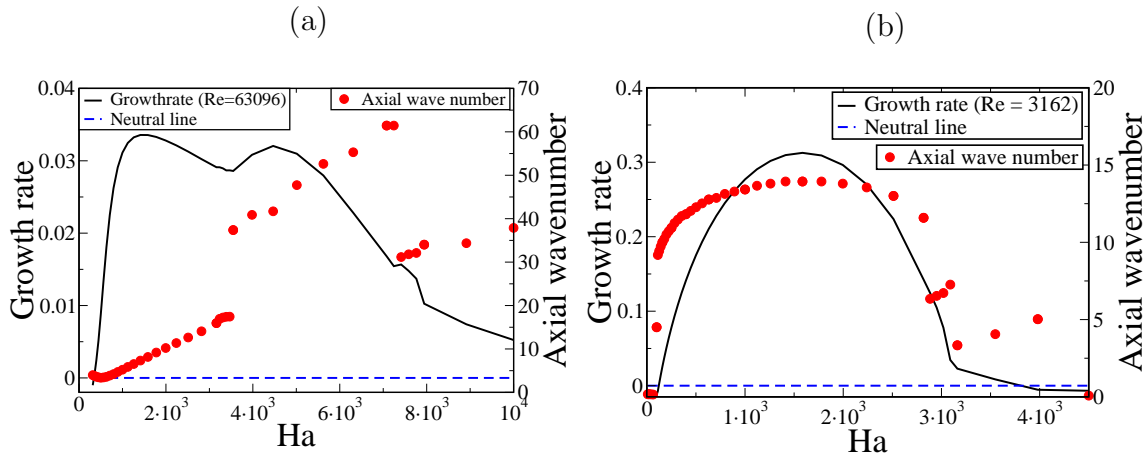}
\put(50,75){\normalsize (b)}
\end{overpic}
}
\caption{Growthrates and axial wavenumbers taken at $Re=const$: ($a$) $Re \approx 6.3 \cdot 10^4$, $Pm=1.4 \cdot 10^{-6}$, ($b$) $Re \approx 3.2 \cdot 10^3$, $Pm=1$. Discontinuous jump in $k$ is accompanied by the change in the slope of growth rate. The blue dashed line corresponds to the neutral curve with $\Re [ \gamma ]=0$ separating linearly stable and unstable regimes.}
\label{Fig:axialwn}
\end{figure}

The instability maps for $Pm=1.4 \cdot 10^{-6}$ ($a$) and $Pm= 1$ ($b$) are presented  in Fig.\ \ref{Fig:stabmaps}. The flow is unstable inside the marked regions. The smaller $Pm$, the more the instability region is shifted upwards, and thus the faster the cylinders have to be rotated to observe instability ($Re_{cr} \sim 10^2$ for high $Pm$ case against $Re_{cr} \sim 10^3$ for low $Pm$). We found that the instability domain is divided into parameter regions where different linear modes of instability dominate. On the borders of these regions the most unstable axial wavenumber $k_{max}$ undergoes discontinuous jumps as $Ha$ is varied (red dots on the Fig.\ \ref{Fig:axialwn}), and, respectively, the growth rate changes slope (solid black lines in Fig.\ \ref{Fig:axialwn}). Recording the parameter values of the jump $(Ha,Re)$, for $Pm=1.4\cdot 10^{-6}$ (Fig.\ \ref{Fig:stabmaps}a) we distinguish the following instabilities: instability I (`basic instability' which appears for both $Pm$, in black), and instability III, arising at high values of $Re$ (yellow line). For $Pm=1$  instability III vanishes and instabilities II (red) and V (blue) arise (see Fig.\ \ref{Fig:stabmaps}b). For both large and small $Pm$ the linear scaling of $Re \sim Ha$ appears at the right border, but at $Pm= 1$ it is extended with instability V (blue region), with scaling of $Re \sim Ha^{0.9}$. 


The green points on Fig.\ \ref{Fig:stabmaps} correspond to the maximum growth rate line. Each point of it marks $Ha$ where the real part of the eigenvalue of the fastest growing mode is maximal (at fixed $Re$). Guseva et al.\ \cite{guseva2015} found that for $Pm=1.4 \cdot 10^{-6}$ this point in the parameter space correlates nicely with the maximum in the torque at the cylinders for fixed $Re$. We follow this parameter path later with direct numerical simulations (brown crosses on Fig.\ \ref{Fig:stabmaps}), estimating an upper bound for transport intensity in this way.

\section{Saturated states of AMRI}

The linear stability analysis does not give any information about the final, saturated state of a system. Fortunately, the critical parameters of $Re \sim 10^3$ and $Ha \sim 10^2$ are accessible in liquid metal experiments. Such experiments indeed have been performed with InGaSn alloy ($Pm=1.4\cdot 10^{-6}$)  \cite{seilmayer2014}. The AMRI was observed as a superposition of two waves, traveling in the opposite direction at slightly different speeds. This was surprising because the azimuthal magnetic field does not break the axial reflection symmetry of a Taylor-Couette system. It was shown numerically in \cite{guseva2015}, that with axially periodic boundary conditions at $Pm=1.4 \cdot 10^{-6}$  the AMRI arises as a standing wave (i.e.  stationary in axial direction), which at the same time rotates azimuthally approximately at the outer cylinder frequency. This wave arises via a supercritical Hopf bifurcation from the laminar flow, where ($Ha-Ha_{cr}$) acts as a bifurcation parameter. However, the standing waves (SW) are stable only close to the onset of instability. At higher $Re$ a subsequent subcritical Hopf bifurcation destabilizes SW and spatial defects accumulate in the system \cite{guseva2015}. More information on symmetries and bifurcation in fluid dynamics in general and Taylor-Couette flow in particular can be found in \cite{crawford1991}. 
\begin{figure}[h!]
\centering
\includegraphics[scale=0.5]{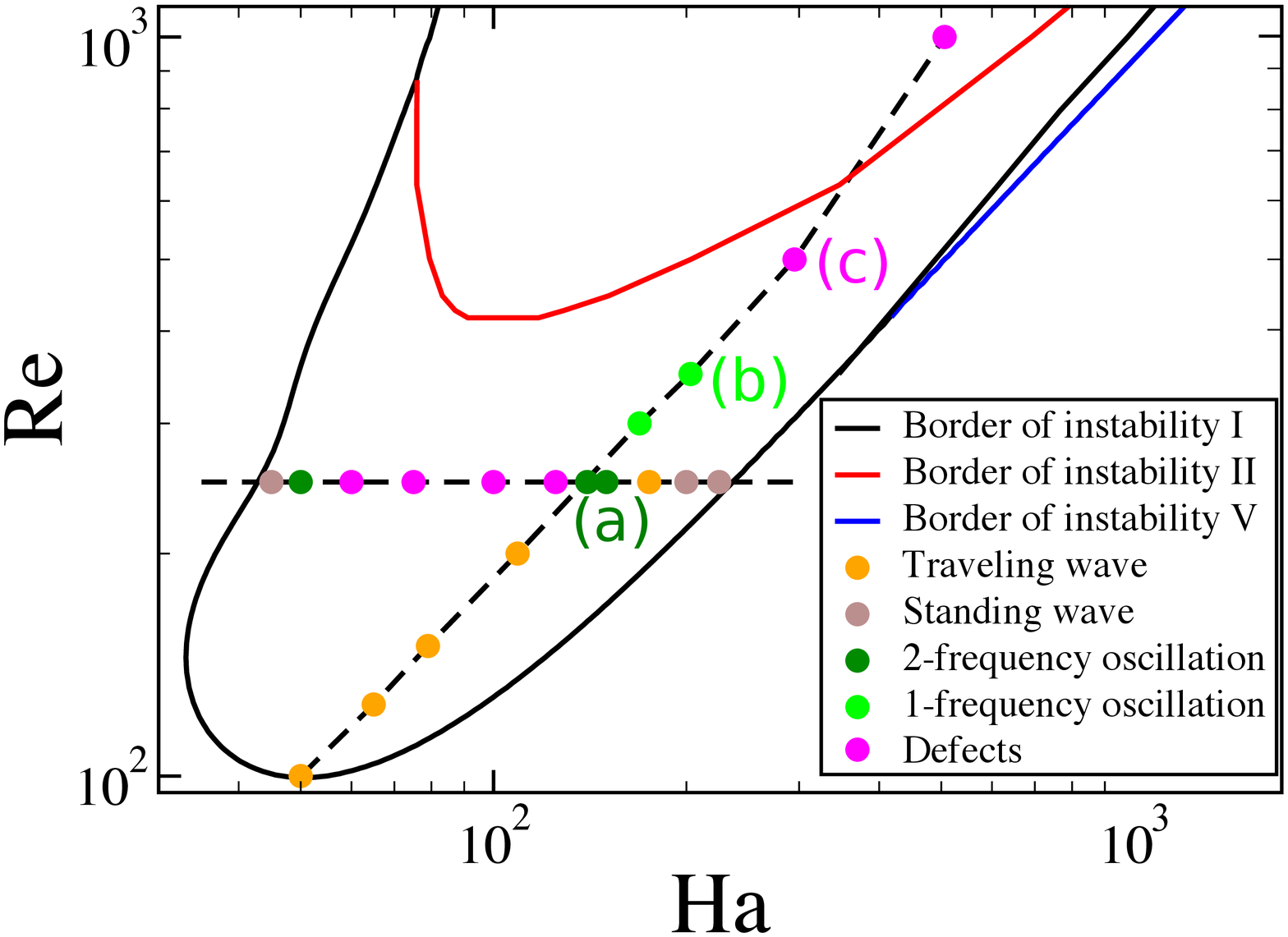} 
\caption{Different flow states found with DNS at $Pm=1$ ($\mu=0.26$): standing waves, traveling waves, 2- and 1-frequency oscillations (of the torque), chaotic solutions (defects). Oscillating in time solutions are stable. The different instability regions are shown as a reference. Vertical dashed line represents the maximum growth rate line, horizontal dashed line - line of $Re=250$. The letters $(a), (b), (c)$ correspond to the states that are shown later on the Fig. \ref{Fig:time_torque} -- \ref{Fig:state_frames}.}
\label{Fig:states}
\end{figure}
%

\begin{figure}[h!]
\centering 
\subfloat[]{
\begin{overpic}[width=0.8\textwidth]{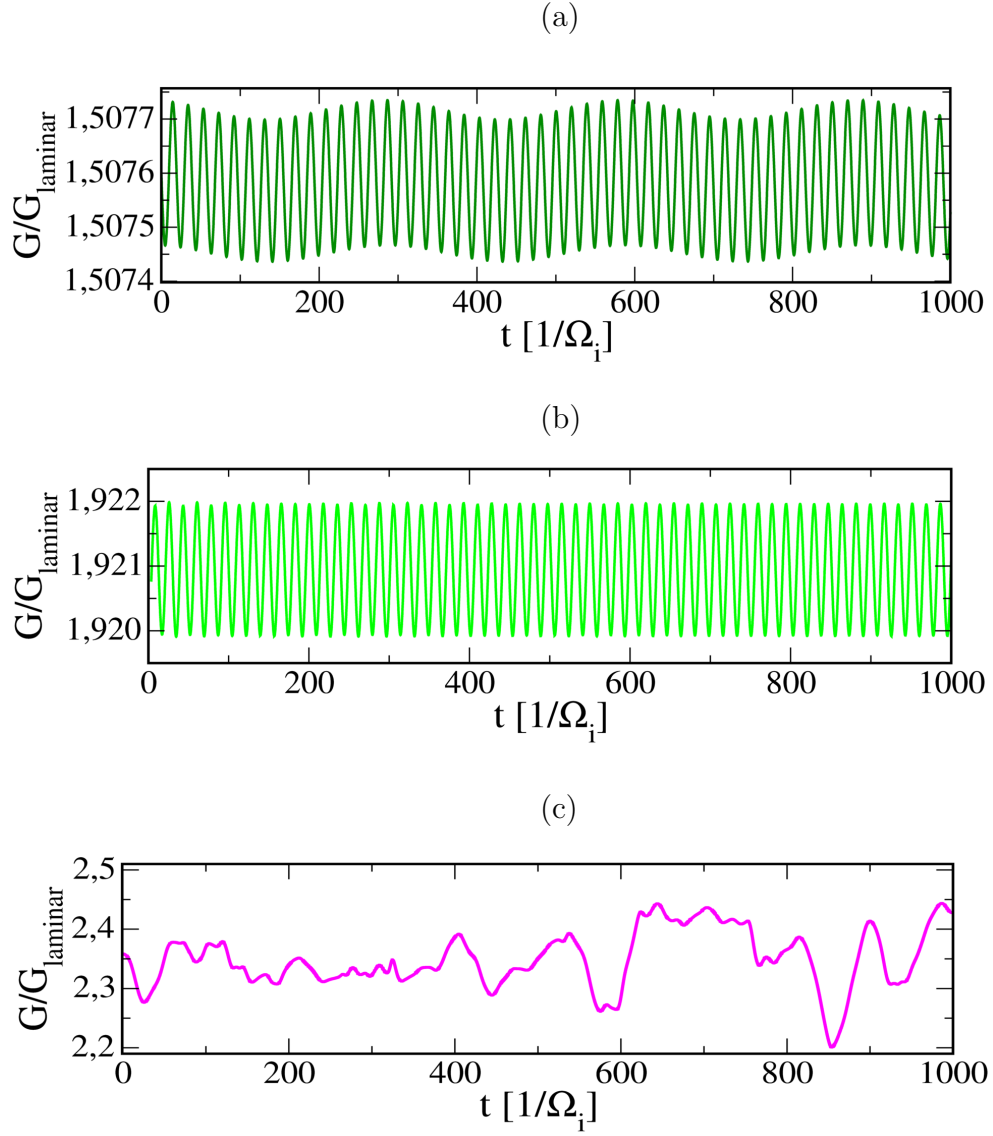}
\put(45,28){\normalsize (c)}
\put(45,61){\normalsize (b)}
\put(45,95){\normalsize (a)}
\end{overpic}
}
\caption{Time series of torque, $Pm=1$: ($a$) 2-frequency oscillation at $Re=250$, $Ha=150$; ($b$) 1-frequency oscillation at $Re=350$, $Ha=203$; ($c$) chaotic solution at $Re=500$, $Ha=295$. The torque is normalized with laminar torque,  time is scaled with rotation period of the inner cylinder $1/\Omega_i$.}
\label{Fig:time_torque}
\end{figure}


\begin{figure}[!ht]
\centering 
\subfloat[]{
\begin{overpic}[width=0.8\textwidth]{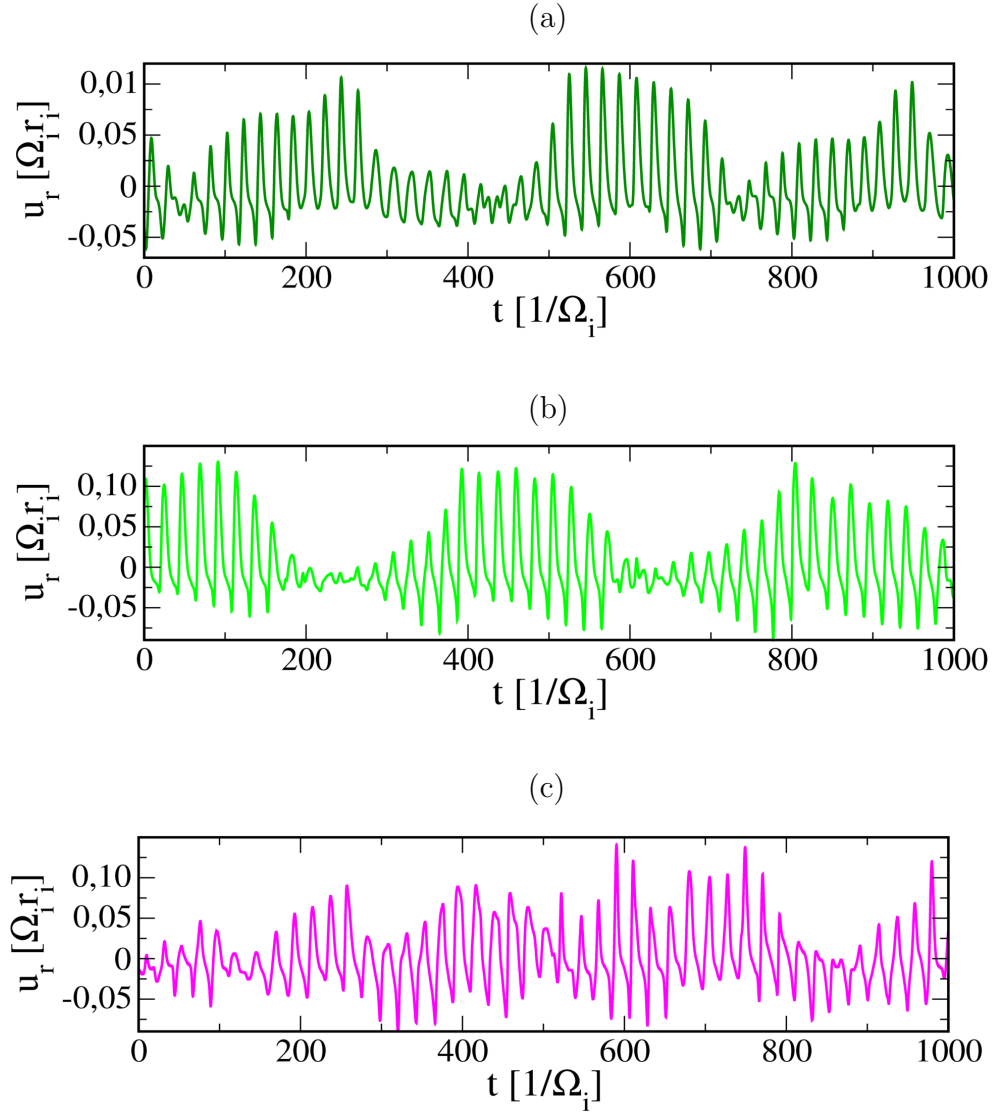}
\put(45,28){\normalsize (c)}
\put(45,61){\normalsize (b)}
\put(45,95){\normalsize (a)}
\end{overpic}
}
\caption{Time series of radial velocity at the point $(r, \phi, z)=(1.5, 0, 0)$, $Pm=1$: ($a$) 2-frequency oscillation at $Re=250$, $Ha=150$; ($b$) 1-frequency oscillation at $Re=350$,  $Ha=203$; ($c$) chaotic solution at $Re=500$, $Ha=295$. Velocity is normalized with the rotation speed of the inner cylinder $\Omega_i r_i$,  time is scaled with rotation period of the inner cylinder $1/\Omega_i$.}
\label{Fig:time_velocity}
\end{figure}

\begin{figure}[h!]
 \begin{tabular}{ccccc}
 (a) & & (b) & & (c)\\ 
\centering
\includegraphics[scale=0.5]{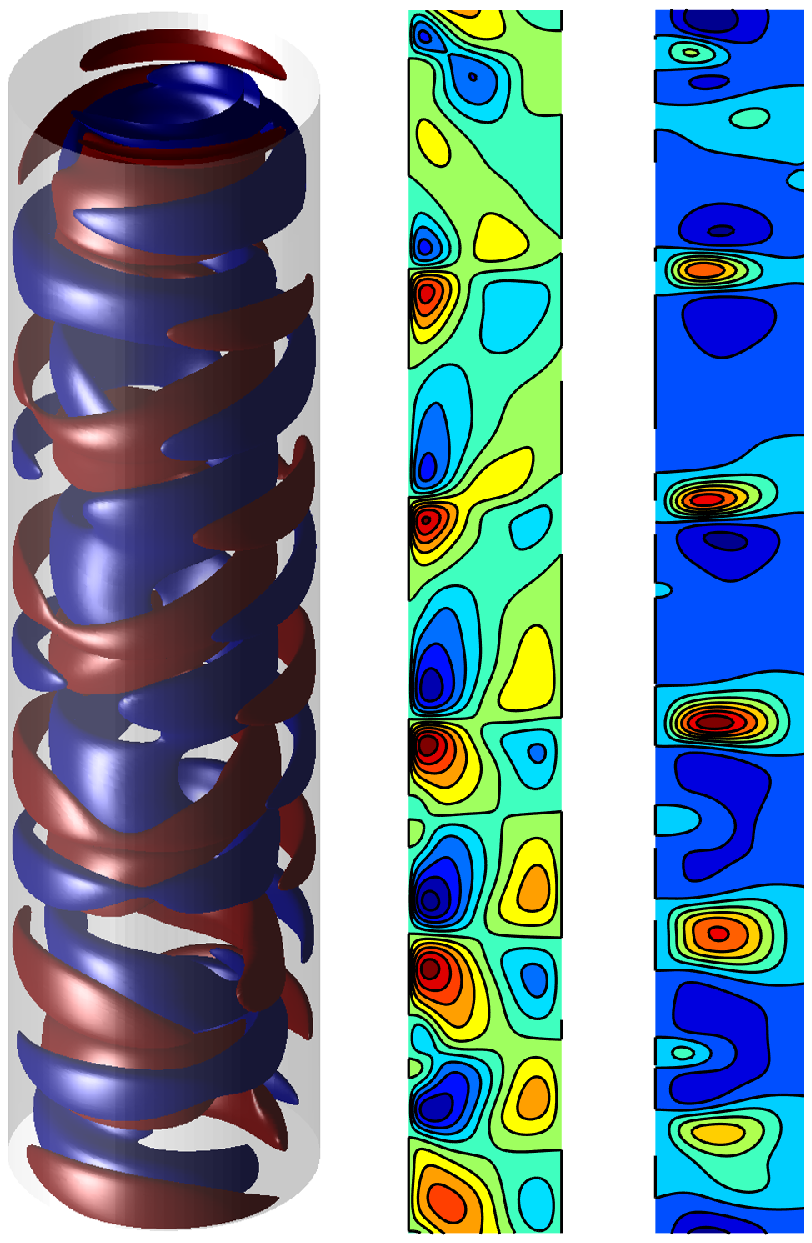} & \hspace{1cm} &
\includegraphics[scale=0.5]{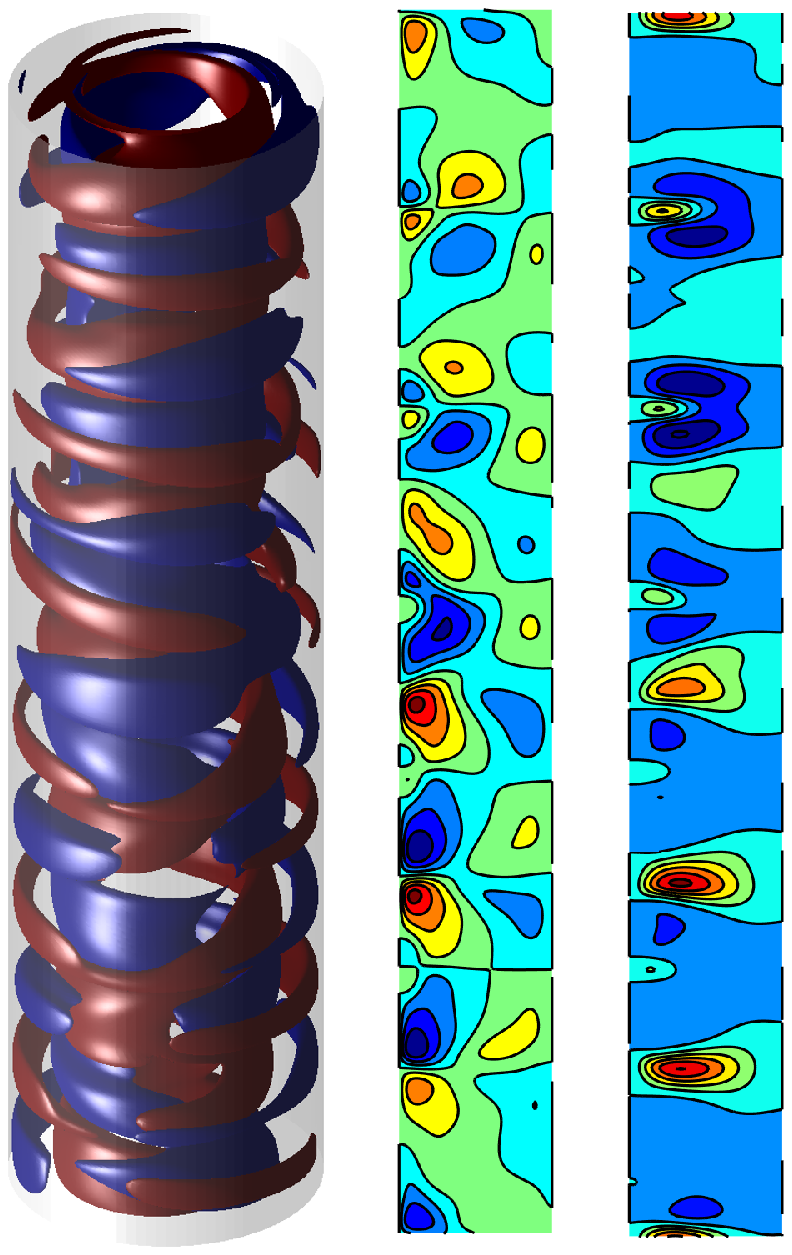} &  \hspace{1cm} &
\includegraphics[scale=0.5]{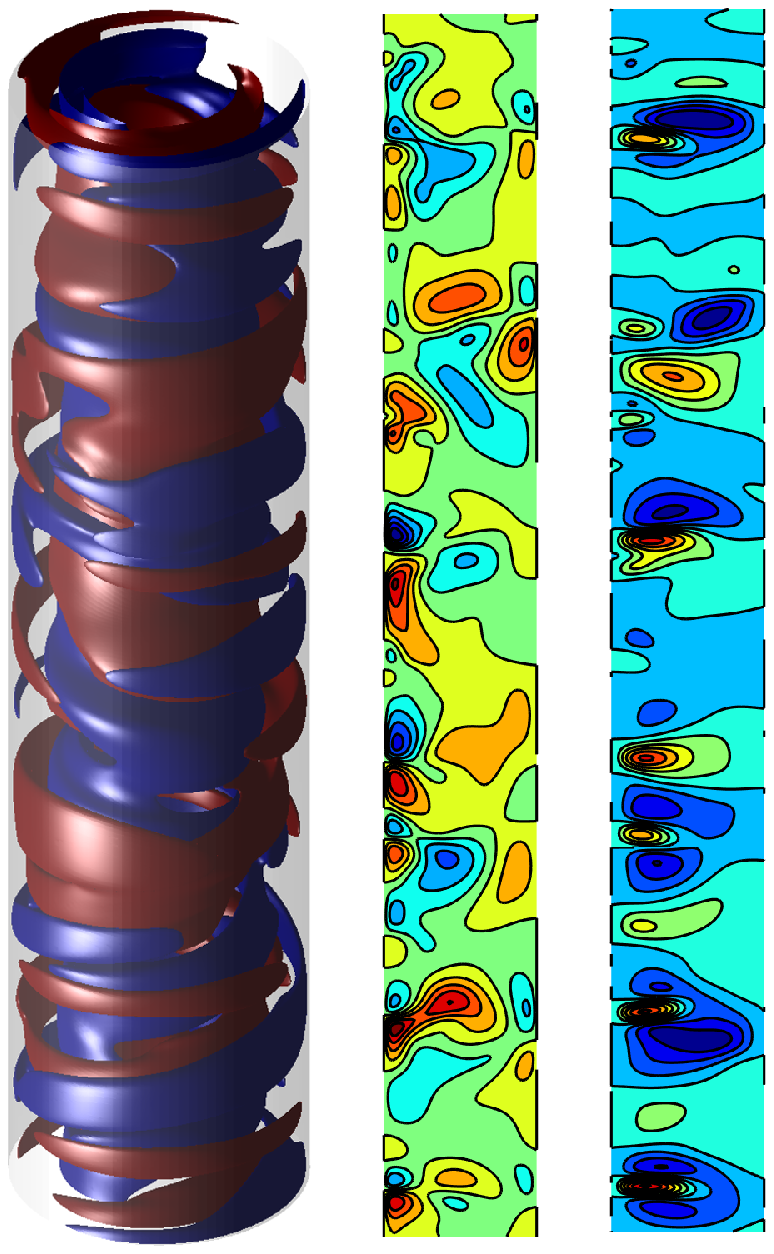}\\
\end{tabular}
\caption{Isosurfaces of axial velocity $\pm u_z$ (left) and contours of axial $u_z$ and radial  $u_r$ velocity (right), $Pm=1$: ($a$) 2-frequency oscillation at $Re=250$, $Ha=150$; ($b$) 1-frequency oscillation at $Re=350$,  $Ha=203$; ($c$) chaotic solution at $Re=500$, $Ha=295$.}
\label{Fig:state_frames}
\end{figure}
The dynamics for $Pm=1$, on which we focus in this paper, seems to be much more rich and diverse than for the small $Pm$ case. The family of different flow states found here is presented in Fig.\ \ref{Fig:states}. The simulations in vertical direction (vertical dashed line) follow the maximum growth rate line. A traveling wave (TW) arises at the onset of instability, followed by 2-frequency solution as $Re$ increases.  The 2-frequency oscillating flow is characterized by modulated oscillations of the torque on the cylinders (see Fig.\ \ref{Fig:time_torque}a), and  the velocity time series have the two frequencies seen in the torque plus the additional one related to rotation of the pattern (Fig.\ \ref{Fig:time_velocity}a). The latter is not present in the torque because it is integrated over all domain and hence is invariant to rotations and translations. Increasing $Re$ further,  we note the transformation of 2-frequency torque oscillation to 1-frequency (\ref{Fig:time_torque}b), with velocity becoming 2-frequency time-periodic (\ref{Fig:time_velocity}b). This is surprising because  more organized flow is more favorable for the system, despite the increase in $Re$. However, if we continue increasing $Re$, the flow becomes chaotic (Fig.\ \ref{Fig:time_torque}c, \ref{Fig:time_velocity}c). The snapshots of the flow in Fig.\ \ref{Fig:state_frames} show that the spatial pattern is complex for both 2-frequency oscillations at $Re=250$, 1-frequency oscillations at $Re=350$, and spatio-temporally chaotic flow at $Re=500$. All cases are characterized by presence of defects which develop on top of the symmetric vortex pattern of the TW at the instability onset.  The 1- and 2-frequency time-periodic solutions also drift axially, similarly to TW. At $Re=500$ the flow is at the onset of turbulence: the vortices of different size are clustered at the inner cylinder and travel up and down with no preferred direction. 

A different scenario is found when $Ha$ is increased and $Re$ is kept constant. The horizontal dashed line of Fig.\ \ref{Fig:states} denotes the simulations that were performed at  $Re=250$. Similar to $Pm=1.4\cdot 10^{-6}$, close to stability boundary the flow is a spatially periodic standing wave (SW), and becomes chaotic in the center of the instability island via a 2-frequency state. A detailed study of bifurcation scenarios is out of the scope of this work  and we leave it as future work.

\section{Angular momentum transport}
\label{sec:dns}
\begin{figure}[h!]
 \begin{tabular}{cc}
 (a) & (b) \\ 
 \centering
\includegraphics[scale=0.3]{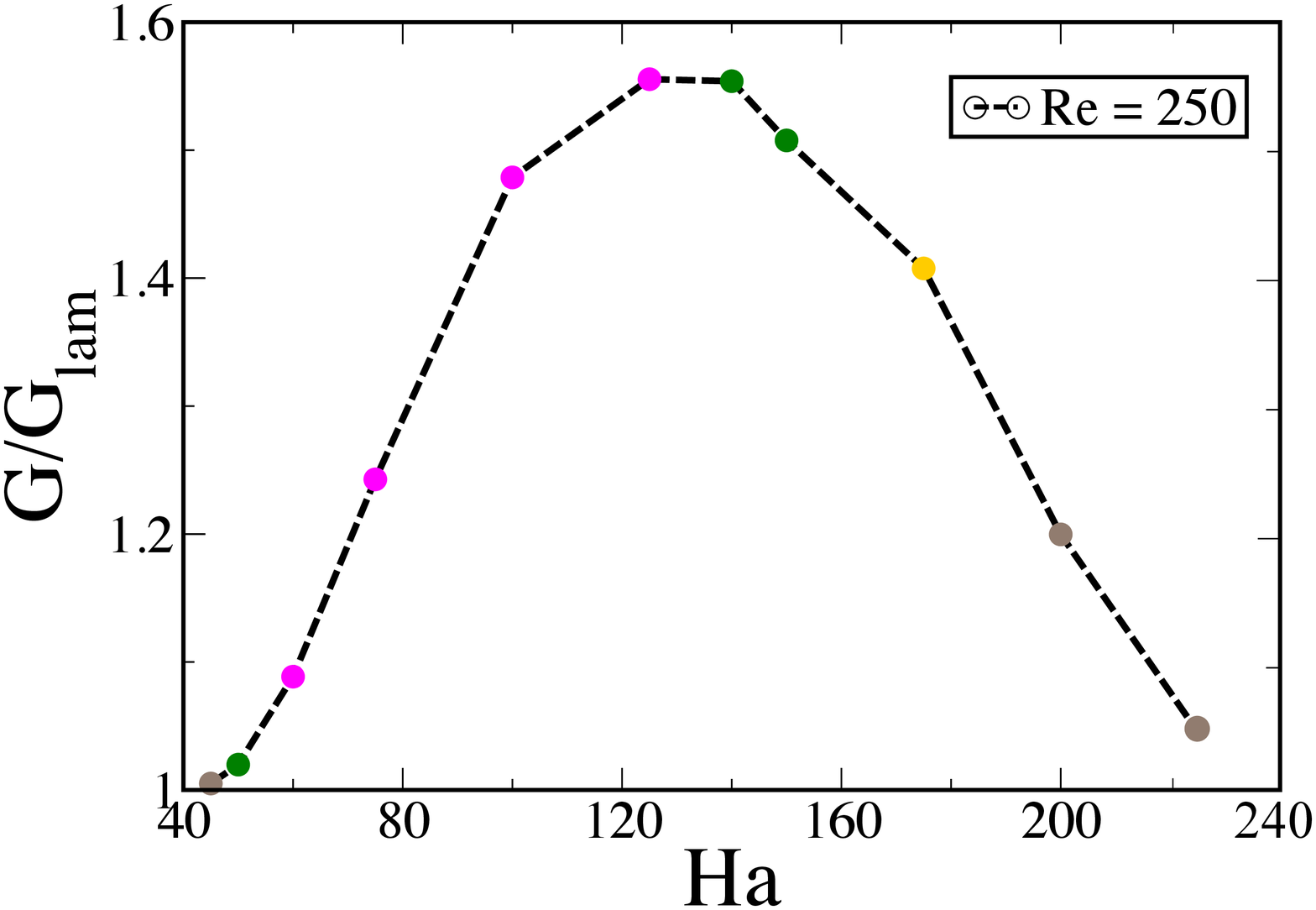} &
\includegraphics[scale=0.3]{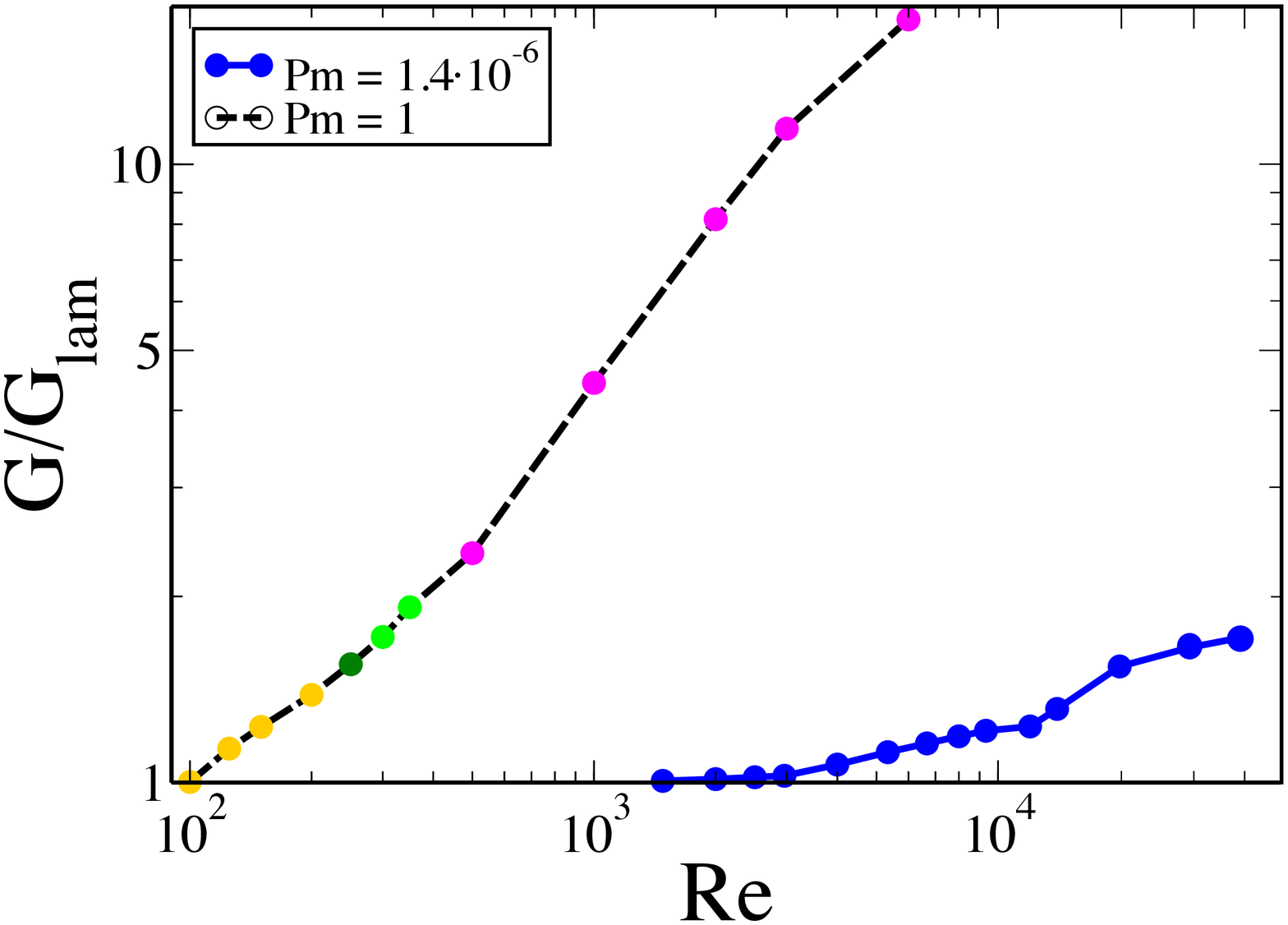} \\
\end{tabular}
\caption{($a$)  $Ha$-dependence of normalized torque at $Pm=1$ along the line $Re=250$;  ($b$) normalized torque enhancement with $Re$ and $Pm$ along maximum growth rate line ($\mu=0.26$). The colored dots in $Pm=1$ represent the different states from Fig.\ \ref{Fig:states}.}
\label{Fig:torque_Pm}
\end{figure}

The transition to turbulence enhances the angular momentum transport, which can be measured as torque at the cylinders. Because of the conservation of angular momentum the average torques on the inner and outer cylinders are equal. Fig.\ \ref{Fig:torque_Pm}a shows the turbulent torque, normalized with the laminar torque, as a function of $Ha$ along $Re=250$ for $Pm=1$. The colored dots correspond to the type of flow observed for $Pm=1$ (see Fig.\ \ref{Fig:states}). The torque, which is small at the onset of instability at low $Ha$, has a maximum around $Ha\approx 130$ and then decreases with further increase of magnetic field.  The flow transits from chaotic pattern back  to 2-frequency solution and then to periodic TW and SW.  The maximum in the torque correlates with the maximum in growth rates ($Ha_{max}=140$), as for $Pm=1.4\cdot 10^{-6}$ \cite{guseva2015}. Fig.\ \ref{Fig:torque_Pm}b shows the turbulent torque as a function of $Re$  and $Pm$ along the maximum growth rate line.   The transition to turbulence at low $Pm = 1.4 \cdot 10^{-6}$ occurs much later than at $Pm=1$: $Re \sim 2 \cdot 10^3$ against $Re \sim 2 \cdot 10^2$. Guseva et al.\ \cite{guseva2015} found that for $Pm=1.4\cdot 10^{-6}$ right after the transition the torque scales with $Re$ as $G \sim Re^{1.15}$. Here we increase the Reynolds number up to $Re=40000$. Our data show that at about $Re=12000$ the torque suddenly changes slope, indicating qualitative changes in the transport properties of the turbulent field. The scaling nevertheless does not change much and can be best approximated as $G \propto Re^{1.16}$, i.e. indistinguishable from the scaling at low $Re$. Hence the transport enhancement at low $Pm$ remains rather small if compared to Taylor-Couette experiments in hydrodynamically unstable regimes \cite{lathrop1992,paoletti2011,vangils2011}. The change of $Pm$ to $1$ influences dramatically the angular momentum transport, which increases by more than an order of magnitude: say, at $Re=6000$ the torque is about $16$ times greater than laminar for $Pm=1$, while at low $Pm$ and the same $Re$ the instability increases transport only by $0.15$. For better understanding of this different behavior of transport quantities it is necessary to study the intermediate values of $Pm$ to find the connections between the two regimes. 


\section{Conclusions}

In this work we studied the magnetorotational instability with applied azimuthal magnetic
field in the Taylor-Couette setup (AMRI). The focus was made on the comparison between two magnetic Prandtl numbers: small $Pm=1.4 \cdot 10^{-6}$ and large $Pm=1$. First, using the method of linear analysis, we found that the right border of instability scales as $Re\sim Ha$ both for high and low $Pm$. For large $Pm$ case the instability is widened up to $Re\sim Ha^{0.9}$, but the line $Re\sim Ha$ remains as a line of discontinuous jump in the axial wavenumber. This jump shows the existence of additional instability modes defining the instability domain.  The types of modes and their regions of existence are different for low and high $Pm$. Second, we followed the linear results with nonlinear direct numerical simulations. The AMRI arises first as a spatially periodic standing or travelling wave, and as the Reynolds number increases the spatially periodic flow turns into turbulence. For low $Pm=1.4 \cdot 10^{-6}$ this transition happens as a sequence of super- and subcritical Hopf bifurcations \cite{guseva2015}, while for $Pm=1$ the transition scenario is much more complicated and involves several spatially complex flow patterns exhibiting 1- or 2-frequency oscillations of integral quantities such as torque or kinetic energy. This phenomenon does not seem to be connected to the existence of several types of modes found in the linear stability analysis of the flow, since it happens at low $Re$ close to the onset of instability and does not touch the lines of discontinuous jump in axial wavenumber $k$, predicted by linear analysis. The competition between several linear modes occurs in the parameter region where the flow is already too turbulent to observe remnants of the linear modes. Following the maximum growth rate lines, we estimated the maximum angular momentum transport efficiency of AMRI up to $Re=6 \cdot 10^3$ for large  $Pm$ and $Re=4 \cdot 10^4$ for low $Pm$. At these values of $Re$ the flow becomes fully turbulent and the turbulent structures do not resemble low $Re$  SW or TW.  As an estimate for the turbulent transport values of torque at the cylinders were computed. Again the small and large $Pm$ showed qualitatively different behavior. For $Pm=1.4 \cdot 10^{-6}$ torque scales as a function of $Re$ as $G \sim Re^{1.15}$. Large $Pm$ flows show much stronger transport enhancement and at $Re=6000$ the turbulent transport is already 16 times larger than the molecular (laminar) transport. These differences in small-$Pm$ and large-$Pm$ flow dynamics demonstrate the need for a study of intermediate $Pm$ for a better understanding of the transition between low- and high-$Pm$ transport properties.

\begin{acknowledgements}
We acknowledge financial support from Deutsche Forschungsgemeinshaft and computing time from Regionales Rechenzentrum Erlangen.
\end{acknowledgements}


\bibliographystyle{ieeetr}






\end{document}